**Title:** Modelling Rayleigh-Bénard convection coupled with electro-vortex flow in liquid metal batteries

**Authors:** Declan Finn Keogh, Victoria Timchenko, John Reizes and Chris Menictas*

**School of Mechanical and Manufacturing Engineering, UNSW, Sydney, NSW 2052, Australia;**

**\*Correspondence:** c.menictas@unsw.edu.au; Tel.: +61-2-9385-6269

**Abstract**

Liquid Metal Batteries (LMBs) are a promising grid-scale energy storage technology that offer low costs per kilowatt-hour, high energy and current densities, as well as low fade rates. The all-liquid composition of the batteries, as well as the presence of temperature gradients and electric and magnetic fields, result in the occurrence of multiple fluid phenomena. These can affect the hydrodynamic stability of the battery, thereby making their interactions critical to understand. In this work, the interaction of Rayleigh-Bénard convection and Electro-vortex flow is investigated as these types of flow will be present in Liquid Metal Batteries from laboratory to grid-scale. A single-layer electrode is simulated, and the computed results compared with experimental data from the literature. It was found that Rayleigh-Bénard convection is unsteady in the liquid metal electrode. The introduction of a 2 A current stabilises the convection cells, whilst the introduction of a 40 A current leads to the dominance of Electro-vortex flow at the central region of the electrode. The results in this work matching experimental data closer than previously published models offering insight into the interaction between Rayleigh-Bénard convection and Electro-vortex flow in the anodes of discharging Liquid Metal Batteries.

1. Introduction

The international community is increasingly looking to renewable energy sources as alternatives to fossil fuels, however, they are generally variable by nature and power generation can be out of synchronisation with demand [1]. To enable the increasing penetration of renewable energy in global electricity grids there is a critical requirement for the storage and release of electricity to align availability of power with demand. Liquid Metal Batteries hold promise as a viable grid-scale storage solution due to their potential for high current densities and low costs per kilowatt-hour [2].

As shown in Figure 1(a), Liquid Metal Batteries consist of three molten layers; an electropositive low density metal that is the anode, an intermediate density salt electrolyte, and an electronegative high density metal that is the cathode [2]. The three layers are immiscible when they are molten and stratified by gravity due to their different densities. The all-liquid construction of the battery is advantageous as it lowers internal resistance in the battery and allows for fast reaction kinetics. Impressive performance characteristics for the most promising chemistries have been demonstrated by different research groups; the Li||Pb-Sb system has been shown to have very low fade rates at 0.004% per cycle [3]; the Li||Te-Sn chemistry can achieve energy densities of 495 Wh/kg and current densities of 3 A/cm$^2$ [4, 5]; and the Li||Bi chemistry has stable discharge voltages at discharge current densities of 1.25A/cm$^2$ and the ability to reverse the formation of intermetallics [6]. This demonstrates that Liquid Metal Batteries have a great potential to enable effective integration of renewables into electricity grids worldwide.

However, the liquid battery technology still has further developmental challenges to overcome before the technology can be utilised effectively at a grid-scale. Of particular importance is the preservation of the integrity of the liquid layers stratified by gravity. The lack of physical barrier between the electrode and electrolyte layers means that the only forces preserving the integrity of the electrode-electrolyte interface is the surface tension and buoyancy forces. Consequently, induced flows of a large-enough magnitude can overcome these stabilising forces, causing the electrode liquids to pierce the electrolyte layer, and come in to contact [7, 8]. The result of this is a short-circuit in which the stored chemical potential energy is released internally in the battery almost instantaneously rather than through the external circuit. Therefore, it is critical for the research area to determine which fluid instabilities occur in Liquid Metal Batteries and their effects on the stability of the battery.

The flows in the three layers in Liquid Metal Batteries are induced by temperature gradients which drive natural convection, and the interaction of the liquid metal electrodes with the electric and magnetic fields in the battery which drive magnetohydrodynamic flows. The streamlines depicted in Figure 1 typify these two types of fluid phenomena. Both the magnitude of the temperature gradients, and the intensity of the electric and magnetic fields increase with the size of battery. This leads to more vigorous flows and creates an upper bound on battery size for safe operation, hereafter called the critical size of the battery [8–11]. For this reason, researchers studying the fluid mechanics of Liquid Metal Batteries have focussed on identifying the convection and magnetohydrodynamic caused instabilities and their effect on the critical size of the battery.

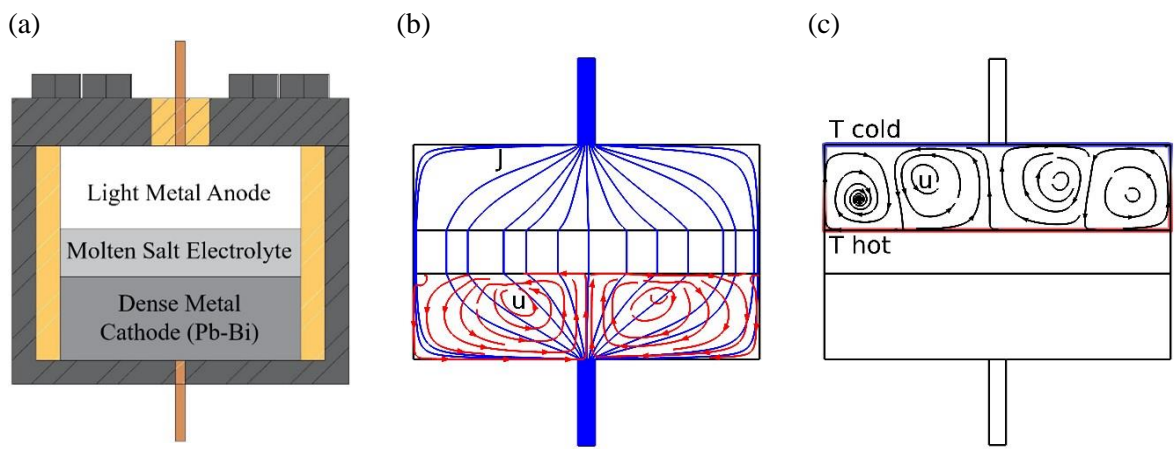

*Figure 1: Flow phenomena in the different layers of liquid metal batteries (a) composition of a liquid metal battery (b) Electro-vortex flow in the cathode (c) convection in the anode.*

The convective instability present in Liquid Metal Batteries is classical Rayleigh-Bénard convection [12]. This instability is characterised by a hot temperature on the lower boundary, and a cold temperature on the top boundary. The magnitude of the difference between the temperatures at the boundaries determines the magnitude of the resultant flow and whether the flow is steady or unsteady. In the batteries themselves, the differences between the boundary temperatures are caused by joule heating from the charge and discharge electric currents [12], heat losses to the environment, as well as electrochemical reactions [13]. The convection regimes in the anode and electrolyte layer are likely to be Rayleigh-Bénard convection [12]. Personnettaz et al. [13] developed both a conduction and convection model of a Li||Bi system and found that a model accounting for heat transfer purely by conduction provided accurate predictions of the temperature profile in the battery when compared to the more complex convection models. Convection has been found to be unsteady and poses no risk to the gravity stratified electrolyte layer [9]. Therefore, Rayleigh-Bénard convection is not considered a brake for the development of the technology.

However, Magnetohydrodynamic instabilities that occur in Liquid Metal Batteries are likely to reduce the critical size considerably. These include the Tayler instability [10, 14–18], the Metal Pad Roll instability[11, 19–24], and Electro-vortex flow [7, 8, 16, 25–27]. The Tayler instability and Metal Pad Roll instability are likely only to affect the critical size of the battery in batteries larger than current prototypes [10] [11], and for this reason they will be excluded from this study. However, Electro-vortex flow has the potential to limit the critical size of Liquid Metal Batteries to lab-scale prototypes namely ~0.1m [8]. Electro-vortex flow is caused by the divergence of the electric current from the vertical axis as it approaches the current collectors. This deviation induces a Lorentz force that leads to a jet-like vortical flow structure as may seen in Figure 1 (b) [16, 18]. This jet can either be poloidal or toroidal depending on the direction of the current supply to the electrode [26]. When the jet is poloidal it can cause both the anode and cathode liquids to push into the electrolyte layer. The resultant pinch-like motion can short-circuit the battery [8].

Electro-vortex flow does not occur in isolation, Rayleigh-Bénard convection can be expected to act simultaneously in discharging batteries. Critically, the interaction of these two flow types and the effects of their combined action on Liquid Metal Batteries has not been thoroughly explored. The first attempts at studying the combined flows were experimental. Traditional fluid visualisation techniques cannot be used in Liquid Metal Batteries due to the high operating temperature and opacity of the working fluid. Instead, Ultrasound Doppler Velocimetry (UDV) measurements of a single component of velocity have been used to characterise the interaction of Electro-vortex flow and Rayleigh-Bénard convection in single layers of molten metal [32–34].

Kelley and Sadoway [28] found that the introduction of Electro-vortex flow into a liquid metal layer at low current densities caused a reduction in spatial variability of the flow as well as an acceleration of the Bénard cells [28]. As the current density increased the magnitude of the flow continued to increase and a fast-oscillatory motion appeared. They concluded that the oscillatory motion could either be due to ohmic heating or traditional magnetoconvection. These observations were reproduced by Ashour et al. in a smaller geometry [30]. While these results have given some insight in to the characteristics of the flow, the data is sparse since only one dimension of velocity in a collimated line was captured. For this reason, numerical simulations offer the greatest potential for characterising the flow.

Currently, there are limited numerical studies in the field that achieve this characterisation- Beltran [31] carried out numerical simulations of the experiment performed by Kelley and Sadoway however, analysis was focussed on the effects of the introduction of Electro-vortex flow on heat transfer in the electrode. Shen and Zikanov [9] numerically studied the effects of introducing a current to a convecting layer and concluded that the resultant flow was only weakly influenced by the current. This contradicts the experimental results of Kelley and Sadoway and Ashour et al. and can be explained by the boundary conditions imposed by Shen and Zikanov. In their work, a homogeneous current was introduced without consideration of the convergence of the current density at the current collectors. This results in a lower peak in the magnetic field in the domain and consequently a smaller Lorentz force. Ashour et al. [30] also attempted numerical modelling of their experiment in their work however, numerical simulations of Rayleigh-Bénard convection and Electro-vortex flow were performed independently [30]. The fact that the simulated data did not match the experimental results is hardly surprising, since the effects of Electro-vortex flow and Rayleigh-Bénard convection need to be modelled simultaneously to capture the effects of the magnetic field on the velocity field.

Considering the characteristic equations of velocity for the two flows, electro-vortex flow should have a higher magnitude of velocity. However, both Ashour et al. [30] and Shen & Zikanov [9] have showed higher velocities induced by convection than electro-vortex flow. This led to Personettaz et al. [13] and Köllner et al. [12]

neglecting the effects of the Lorentz force induced flow altogether in their model. Considering that Rayleigh-Bénard convection is not expected to cause electrolyte layer rupture for Liquid Metal Batteries while Electro-vortex flow could cause rupture in batteries with a radius as small as 0.1 m it is critical to determine which flow can be expected to dominate in charging and discharging batteries. The work of Ashour et al. [30] represents the most recent experimental results so it has been selected as a basis for the numerical model developed in this work. In the following sections sufficient details of the experiment performed by Ashour et al. [30] will be presented to allow the numerical model to be built.

2. **Numerical Model**

To reproduce the experimental set-up of Ashour et al., the computational model was built with the geometry shown in Figure 2. The experiment was carried out with a single layer of eutectic Pb-Bi alloy. While Rayleigh-Bénard convection and Electro-vortex flow are flows typical of an anode, a cathode alloy was used due to the much lower melting temperature. The temperature dependent material properties of eutectic Pb-Bi were sourced from [36–38]. The material properties required for the simulation can be found in Table 1.

*Table 1: Material properties used to perform the numerical simulations*

| Material | Property | Value |
|---|---|---|
| Eutectic Pb-Bi at 156°C | Kinematic viscosity $v$ *($m^2/s$)* | 2.724E-07 |
| | Thermal volumetric expansion coefficient $\beta$ *(1/K)* | 1.257E-04 |
| | Electrical conductivity $\sigma$ *(S/m)* | 896,867 |
| | Density $\rho$ *($kg/m^3$)* | 10,510 |
| | Isobaric heat capacity $c_p$ *(J/kgK)* | 147.7 |
| | Thermal conductivity $\lambda$ *(W/mK)* | 9.8 |
| | Permeability $\mu_0$ *(H/m)* | 1.25664E-06 |
| Copper current collector at 151°C | Electrical conductivity $\sigma$ *(S/m)* | 3.90E+07 |
| | Permeability $\mu_0$ *(H/m)* | 1.25664E-06 |
| Stainless steel 304 vessel | Electrical conductivity $\sigma$ *(S/m)* | 1.37E+06 |
| | Permeability $\mu_0$ *(H/m)* | 1.26129E-06 |
| Aluminium block | Electrical conductivity $\sigma$ *(S/m)* | 3.70E+07 |
| | Permeability $\mu_0$ *(H/m)* | 1.25667E-06 |

Since the model being developed is a magnetohydrodynamic model the magnetic Reynold's number must be considered before deciding on the appropriate formulation of the equations. For magnetic Reynolds numbers much less than one, the magnetic field will remain unperturbed by the velocity field. When the magnetic Reynolds number is much greater than one, any inhomogeneity induced by the velocity field no longer diffuses away, making it necessary to re-calculate the magnetic field at each timestep [35]. The magnetic Reynolds number is given by

$$R_m = \sigma \mu_0 U_{max} l \qquad (1)$$

and can be calculated by using the expected maximum fluid velocity. Considering the work of Ashour et al. [30] the max velocity of the fluid is not expected to exceed ~30mm/s. Taking the electrical conductivity of PbBi at 156°C, 896867 S/m, the vacuum permeability, and the characteristic length of the geometry $l = 0.011$ m, $R_m$ is found to be $3.7 \times 10^{-7}$. Accordingly, the low magnetic Reynold's number approximation is taken as described by P.A. Davidson [35]. In this approximation the magnetic field is a summation of the externally applied magnetic field and the induced magnetic field so that

$$\mathbf{B} = \mathbf{B_e} + \mathbf{B_\theta} \tag{2}$$

in which $\mathbf{B_e}$ and $\mathbf{B_\theta}$ are the induced and applied magnetic field, respectively. The induced magnetic field is generated by the electric current supplied to the electrode and the applied magnetic field is the Earth's magnetic field. The coupling of the magnetic field to the velocity field is not considered and consequently the field is static throughout the calculations.

Similar to the magnetic field, the electric potential and current density are considered to be compromised of static and induced components

$$\phi = \phi_e + \phi_i \tag{3}$$

$$\mathbf{J} = \mathbf{J}_e + \mathbf{J}_i \tag{4}$$

in which $\phi_e$ is the applied potential, $\phi_i$ is the induced potential, $J_e$ is the applied current density, and $J_i$ is the induced current density as in the work of Ashour et al. [30]. To determine the applied component of the electric field the laplace equation for electric potential

$$\nabla \sigma \nabla \phi_e = 0 \tag{5}$$

is solved in which $\sigma$ is the electrical conductivity of the medium. To account for the changes in conductivity between the different regions the numerical scheme developed by Weber at al. in [27] is implemented.

Once the applied potential has been computed the magnetic vector potential can be determined

$$\nabla^2 \mathbf{A}_e = \frac{1}{\sigma\mu} \nabla \phi_e \tag{6}$$

in which $A_e$ is induced magnetic vector potential from the electrical field and $\mu$ is the permeability of the medium. The induced magnetic flux intensity and applied current

$$\mathbf{B}_e = \nabla \times \mathbf{A}_e \tag{7}$$

$$\mathbf{J}_e = -\sigma \nabla \phi_e \tag{8}$$

are found to complete the computation of the static components of the electromagnetic fields. Next the induced components of the fields are calculated. The induced electric potential from the velocity field is calculated using

$$\nabla \sigma \nabla \phi_i = \nabla \cdot \sigma(\mathbf{u} \times \mathbf{B}) \tag{9}$$

And the induced current density is found using

$$\mathbf{J}_i = \sigma(-\nabla \phi_i + \mathbf{u} \times \mathbf{B}) \tag{10}$$

in which $u$ is the three-dimensional velocity field. The electric and magnetic fields are coupled to the conservation equations of momentum, also known as the Navier-Stokes, continuity, and energy equations through the calculation of source terms for joule heating

$$Q_J = \frac{J^2}{\sigma} \tag{11}$$

And the Lorentz force

$$\mathbf{f_L} = \mathbf{J} \times \mathbf{B} \tag{12}$$

These source terms are then included in the incompressible formulation of the Navier-Stokes, continuity, and energy equations

$$\frac{\partial \mathbf{u}}{\partial t} + (\mathbf{u} \cdot \nabla)\mathbf{u} - \nu \nabla^2 \mathbf{u} = -\frac{\nabla p}{\rho_0} + \frac{\mathbf{f_L}}{\rho_0} - \frac{1}{\rho_0}\mathbf{f_T} \qquad (13)$$

$$\nabla \cdot \mathbf{u} = 0 \qquad (14)$$

and

$$\frac{\partial T}{\partial t} + (\mathbf{u} \cdot \nabla)T = \frac{\lambda}{\rho_0 C_p}\nabla^2(T) + \frac{1}{\rho_0 C_p}Q_J \qquad (15)$$

The Boussinesq-Overbeck approximation was used in which it is assumed that all the transport properties are constant including density except in the buoyancy term of the momentum equation, in which it is assumed to be a linear function of the temperature. Finally, this buoyancy term is given by

$$\mathbf{f_T} = \Delta\rho \mathbf{g} h \qquad (16)$$

where

$$\rho = \rho_0\big(1 - \beta(T - T_0)\big) \qquad (17)$$

Equations (4-17) were solved using a custom solver implemented in OpenFOAM, an open source finite volume library. As a first step, grid convergence was tested with by averaging the instantaneous volumetric mean velocity over 100s and comparing the values for differently sized grids. There was a 5% difference in the value of the velocity averaged over space and time between a mesh with 563,618 cells and a mesh with 6,290,952 cells, but only a 1% difference when the grid was refined to have 10,488,960 cells. Therefore, a grid with 6,290,952 cells was utilised for the simulations.

### 3. Geometry and Boundary Conditions

There were two geometries used to perform the numerical calculations and they can be seen in Figure 2; both geometries match the experimental setup of Ashour et al. [30]. The first geometry was used for calculations of the background electric and magnetic and the result was then mapped on to the fluid domain. The details of the experiment are briefly summarised below, as well as the boundary conditions utilised in the simulations. For greater detail refer to [28–30].

The Pb-Bi electrode was contained in a vessel made from stainless-steel 304. The vessel had an inner diameter, ∅, of 88.9 mm, an outer diameter of 100 mm, and was heated from below and insulated on the top and sides. The vessel was placed on an aluminium block through which a current was supplied, allowed to pass through the molten Pb-Bi layer, and collected using a ∅4 mm nickel-plated copper wire. In the simulations the entire domain was captured when solving for the electric and magnetic field, including the stainless-steel vessel, current collector, and aluminium block. The results were then mapped on to a mesh of solely the liquid Pb-Bi region. For computations of the magnetic field an external magnetic field equivalent to the earth's magnetic field measured in Dresden, $\mathbf{B_\theta} = (15 \cdot \mathbf{e_x}, 5 \cdot \mathbf{e_y}, 36 \cdot \mathbf{e_z})$ micro Tesla ($\mu T$) was applied to all surfaces.

The base of the vessel was maintained at a temperature of approximately 160°C using a PID controller. A temperature gradient of 7-9K formed between the top and the bottom of the liquid layer due to heat generation and dissipation. As in the work of Ashour et al. [30] this temperature gradient was created using Dirichlet boundary conditions for temperature. The base was set to 160°C, the top set to 152°C, and the surface where the current collector touches the liquid metal layer set to 151°C. The sidewalls were assumed to be adiabatic since the sidewalls of stainless-steel vessel were insulated. A UDV probe, immersed in the electrode through a port, captured and averaged the horizontal velocity in cross-sectional cuts. The profile of the beam for the transducer used by Ashour et al. [30] can be found on the sound processing website [36]. All liquid boundaries were assumed to be no-slip impermeable walls. This includes the Argon-PbBi free-boundary since the presence of an oxide layer on top of the electrode [30], [37] prevents motion and mixing.

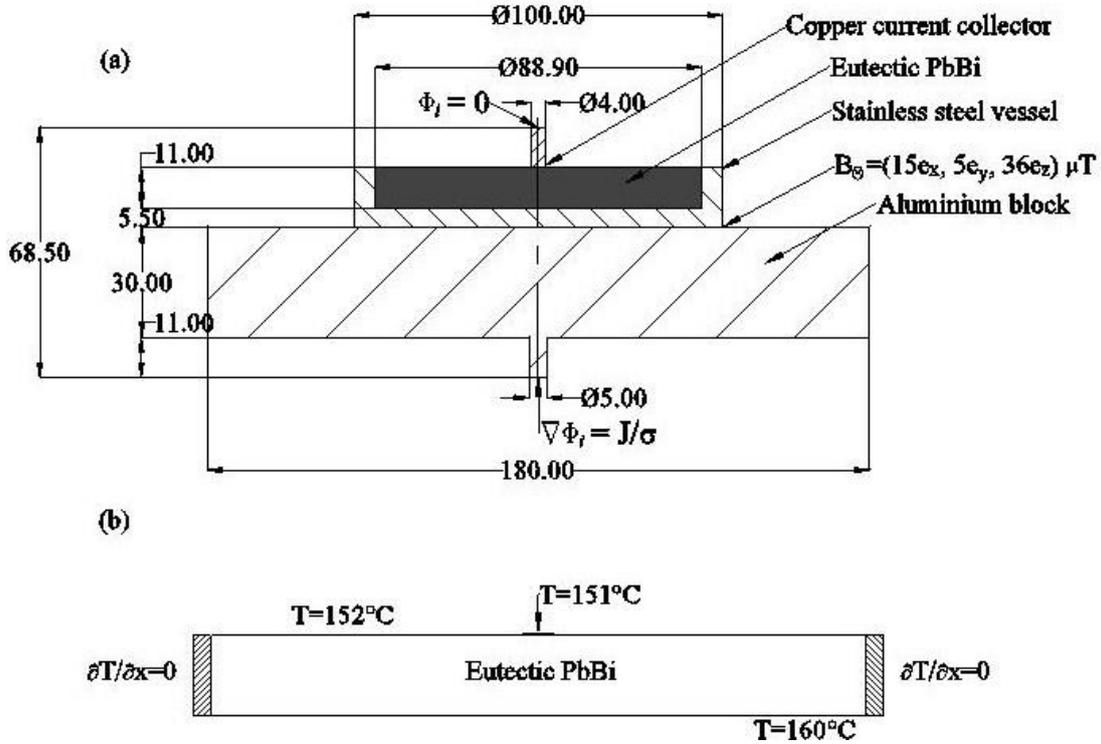

*Figure 2: Dimensions (in mm) and boundary conditions of the three-dimensional numerical geometry used in the simulations (a) the geometry used for calculations of the background magnetic and electric field (b) fluid domain*

### 4. Results

In Ashour et al.'s [30] work, the effects of Rayleigh-Bénard convection and Electro-vortex flow were distinguished by first recording data for the case without current. Then, subsequently, data for the cases with 2 A and 40 A current were recorded. For this reason, the current work will be presented in the same order. In all cases Equations (2)-(17) were solved. The convection simulation at t=300s was used as the initial condition for the 2 A simulation. Similarly, the fields from the 2 A simulation at t=300s were used as initial conditions for the 40A simulations. This is an important consideration as convection can exhibit hysteresis [38].

#### 4.1 Convection case

In the initial simulation without a current applied the convective pattern in the electrode initially consisted of four concentric annuli that create a target pattern. However, after t≅50s the system diverges from this stable pattern. Indeed, the velocity patterns on the mid plane of the liquid layer change continuously with the shapes shown in Figures 3 (a and b) being two instantaneous snapshots. Similarly, the convective cells are also transient in all three-dimensions. Compared to fluids with Prandtl numbers (Pr) > 1, fluids with Pr < 1 have a lower critical Rayleigh numbers at which the flow becomes unsteady in Rayleigh-Bénard convection [39]–[42] and so the unsteadiness of the flow is to be expected of a fluid with a Prandtl number as low as that of the Pb-Bi alloy, viz, Pr = 0.04. This finding is consistent with previous studies on convection in Liquid Metal Batteries [9], [12].

Since the flow is unsteady, the mean flow is used to establish its general characteristics. Further, this allows comparisons of the mean experimental values presented by Ashour et al. [30]. As discussed in Section 3, the experimental data was recorded with a UDV probe. The probe measured the horizontal component of velocity, with flow towards the transducer having a negative value. Since the midline of the probe was situated above the mid-plane of the flow, looking from the negative axis to the positive, the negative values recorded by the probe represent Bénard cells rotating counter-clockwise while positive values represent Bénard cells rotating clockwise. For a more in-depth description of the workings of UDV, refer to the work of Kelley and Sadoway [28] and Perez and Kelley [32].

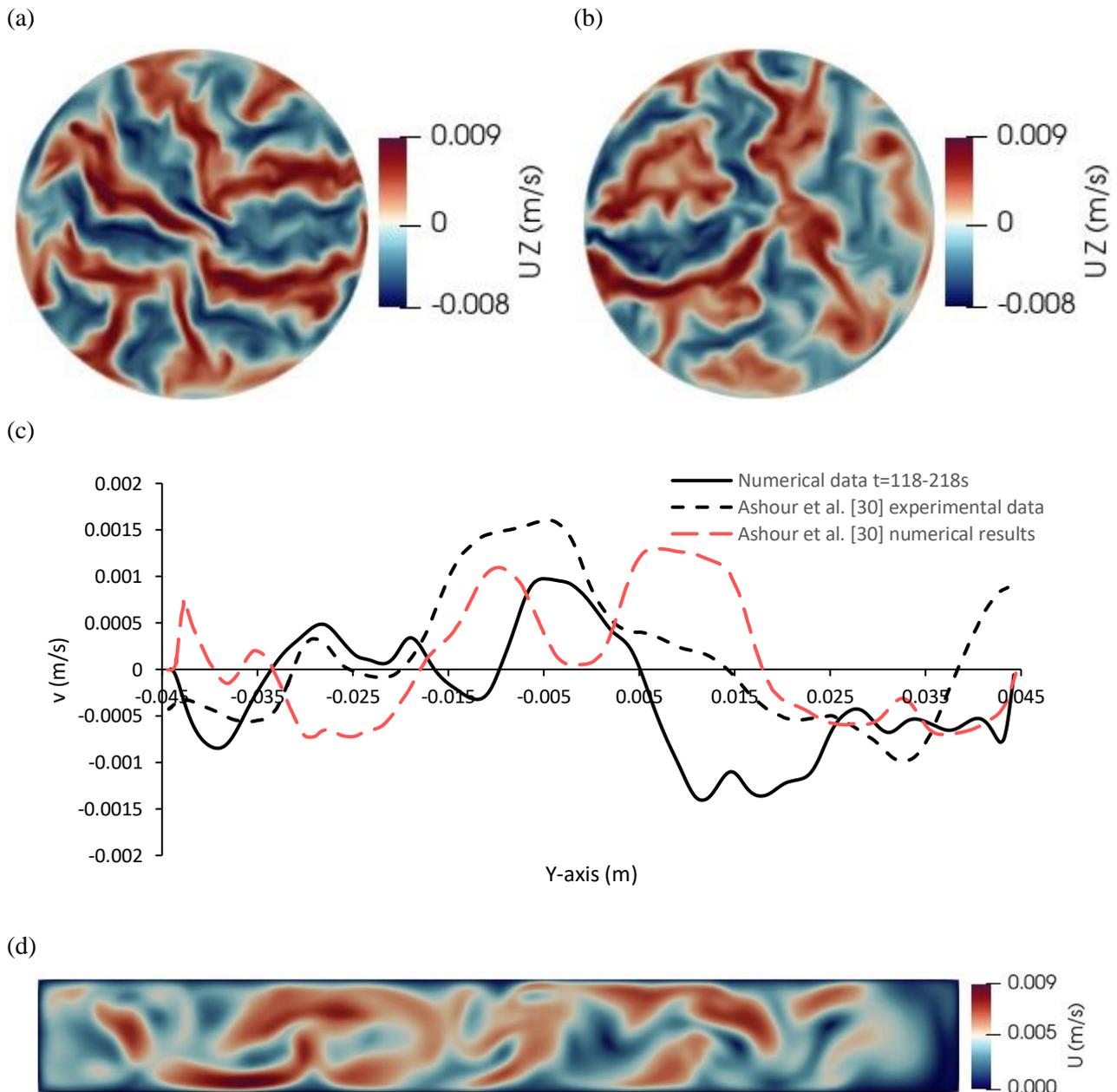

*Figure 3: Convection in a Liquid metal electrode (a) and (b) instantaneous UZ-velocity contour plots of the xy mid-plane at t=120s and t=170s respectively (c) Comparison of the mean velocity profile averaged between t=118 and t=218s and experimentally obtained in [30] (d) instantaneous velocity magnitude contour plot on the yz-plane at t=170s*

Starting on the left-hand side of the computed UDV probe plot in Figure 3 (c), the time averaged velocity data show an anti-clockwise rotating cell adjacent to the sidewall, two smaller cells following this, a large clockwise rotating cell under the current collector, and a single large anti-clockwise cell to the right of the current collector. Comparing between the computed results of this work and experimental data of Ashour et al. [30] it may be observed that the number of rolls, (five), and peak magnitude of the average velocities match closely at $v \cong 0.0015 \, m/s$. It can be also seen that there is not a substantial difference between the numerical results presented in this work and that of Ashour et al. [30] except that the peak magnitude of the velocity data obtained in this work more closely matches the experiment. However, the maximum time-averaged velocity in the simulations is on the opposite side of the origin. If the average of a different set of times is used to calculate the profile its shape changes due to the transient nature of the convective instability. It seems that the key to matching the profile could be in finding the correct set of times steps to compile the average value, however the times used in the experiment were not reported. Another explanation for the differences between the experimental and simulation data is that the boundary conditions used in the computational model may not represent the experimental boundary conditions perfectly. Ashour et al. [30] identified that the temperature of the top surface of the liquid metal layer varied by as much as two degrees throughout the experiment. They also suggested that the side walls might not be perfectly insulated. This could create a substantial difference in the resultant convection pattern and velocity. Timchenko and Reizes [43] have demonstrated the extreme sensitivity of natural convection to boundary conditions and the difficulty this presents when validating numerical models and so the differences between the experiment and computation results do not detract from the validity of the simulations.

### 4.2 2 A case

Next, the effects of a weak Electro-vortex flow on Rayleigh-Bénard convection are considered through the introduction of 2 A current to the layer. As described in Section 3 the current is supplied through an aluminium block and allowed to pass through the layer before being collected at the axially located current collector. The results are then mapped on to a separate mesh of the liquid layer. There is no immediately visible effect upon the introduction of the current. This is because the electro-vortex flow is significantly weaker than the Rayleigh-Bénard convection, and so the characteristics of the flow are primarily determined by the temperature gradient, not the introduced current.

However, in Figure 4 (a) and (b) the vertical velocity of the convection rolls has decreased from 9 mm/s to 8 mm/s. Also, from qualitatively comparing Figure 3 (d) and Figure 4 (c) the convection cells have become steadier after the introduction of the current. The standard deviation across the probe sampled data was used to quantitatively determine whether the characteristics of the flow had been altered.

(a) (b)

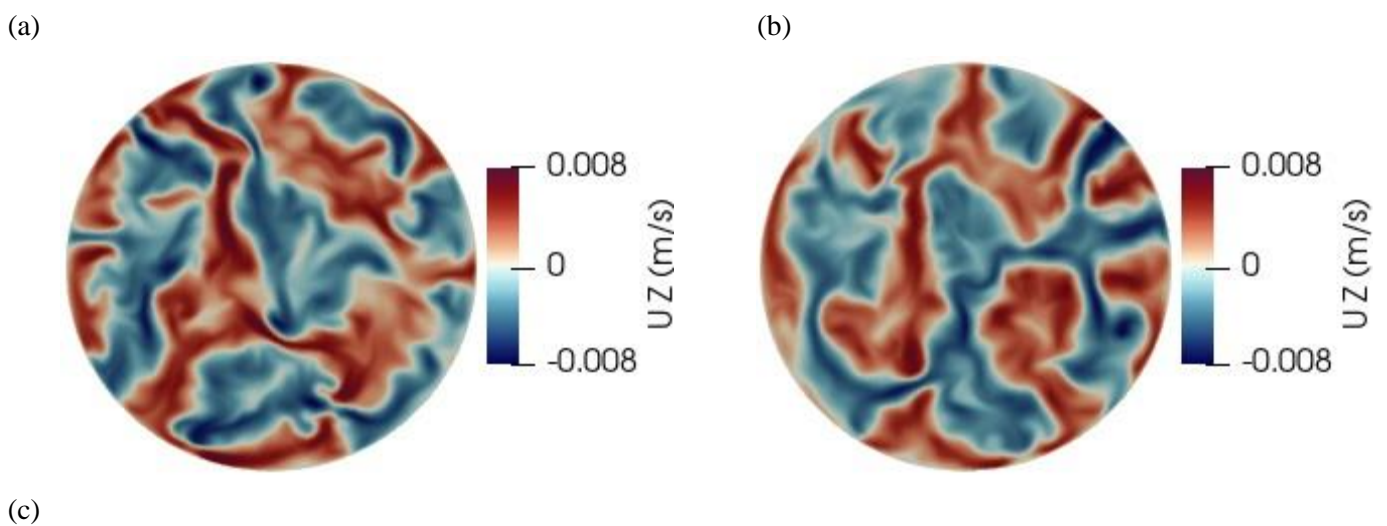

(c)

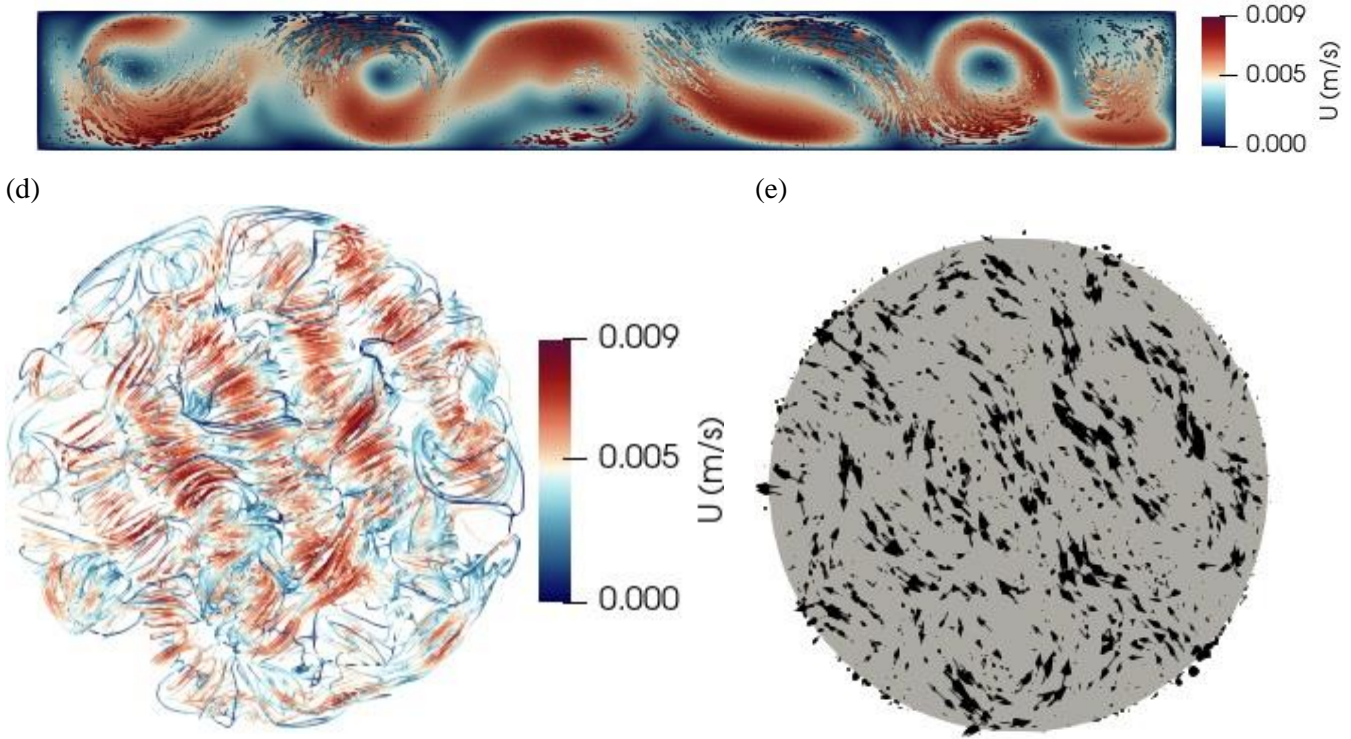

*Figure 4: Convection with 2A of current: instantaneous velocity contour plot of the UZ component on the xy mid-plane at (a) t=224s and (b) 274s; (c) convection cells on the xz midplane at t=154s (d) velocity streamlines showing the convection cells at t=154s & (d) the induced current density vectors at t=154s*

The equation for sample standard deviation is given by

$$s = \sqrt{\frac{\sum(x_i - \bar{x})^2}{n-1}} \qquad (17)$$

in which n is the size of the sample, $x_i$ is the instantaneous values of vertical velocity, $\bar{x}$ is the mean velocity, and $s$ is the standard deviation. Samples were taken at 0.05 s intervals and, during the times sampled, it was found that for the convection case the standard deviation was 0.00169 while for the 2A case the standard deviation was 0.00162. This supports the qualitative observations that the flow has become steadier upon the introduction of 2 A current. This result is different from that of Shen & Zikanov [9] who found no evidence of magnetohydrodynamic effects on convection in Liquid Metal Batteries. Shen & Zikanov [9] modelled a three-layer liquid metal battery with a current supplied uniformly to the cathode and collected uniformly at the anode. The use of a homogenous current density at the anode instead of a point-like current collector is unrealistic and does not reflect the way that current is collected at the anode in present designs. Therefore the steadying of the convection by application of 2 A of current can easily be explained by the significant convergence of the current density at the current collector in the present model. The high current density at the current collector increases the induced magnetic field. The increased magnetic field induces a current that deaccelerates the convection rolls and steadies the flow field. Figure 4 (d) presents a streamline plot of the convection rolls while Figure 4 (e) shows the vectors of the induced current. The direction of the induced current along the convection cells is perpendicular to the direction of the velocity causing a Lorentz force opposing the cells. Though this Lorentz force is small in magnitude, it is large enough to make the convection cells steadier.

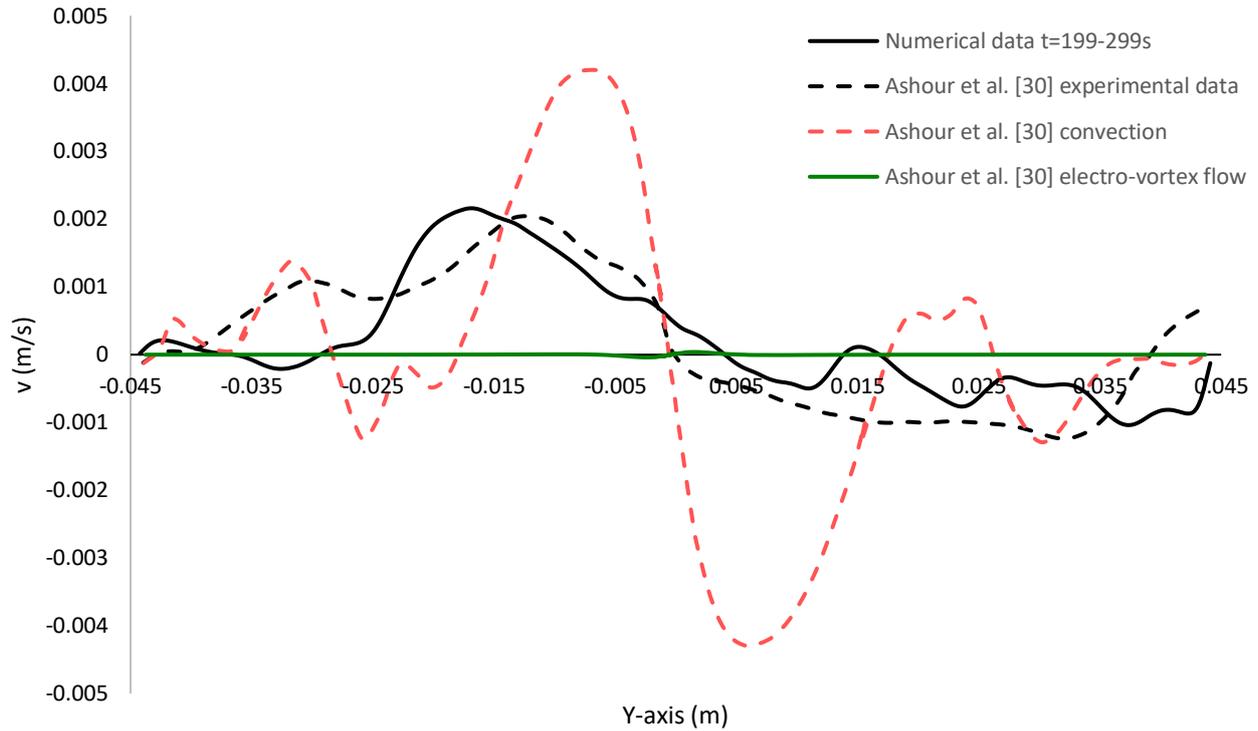

*Figure 5: Comparison of the numerically predicted mean velocity profile for t=199-299s and the experimentally obtained profile*

The numerically generated mean profile from the times t=199-299s and the experimental profile from Ashour et al. [30] are plotted in Figure 5. It is clear that the numerical predictions for the flow obtained in this work closely agree with the experimental data both in magnitude and shape, however, there are still discrepancies related to the behaviour of the convection cells. Since at this current density the Rayleigh-Bénard convection is dominant, similar to the convection case, the differences observed are caused either by a compilation of different timesteps to the experiment when obtaining the average velocity values or implementation of boundary conditions that do not exactly reproduce the experiment. Despite this, like in the convection case, the behaviour of the system for the 2 A case has been captured by the current numerical model reasonably well. A much higher degree of accuracy of the current model which includes coupling of Rayleigh-Bénard convection and electro-vortex flow is also confirmed by comparison with numerical results of Ashour et al. [30] where these effects were separated as shown in Figure5.

A large upwards flow from joule heating under the top current collector is not present in this numerical result in contrast to Ashour et al. [30] leading to the conclusion that the alterations to the flow are caused by the electro-vortex flow. Therefore, it can be concluded that even a weak electro-vortex can impact Rayleigh-Bénard convection in Liquid Metal Batteries and that electro-vortex flow at small current densities will stabilise convection in the electrochemical cells.

### 4.3 40 A case

An increased 40 A of current was applied to the electrode with initial conditions supplied by the 2 A case. Unlike the 2A case, an electro-vortex jet is evident from the time of application of the 40A current. Figure 5 (a) shows the jet induced by the electromagnetic force dominating in the central region of the liquid layer, while convection dominates outside of the central region. This localised dominance of flow types is similar to the interaction of buoyancy and Lorentz force presented by Davidson et al. [44]. However, there is still considerable interaction of the two flow types. As can be seen in Figure 6 (a) and (b) the electro-vortex jet is swept off the vertical axis by the adjacent convection rolls and the jet oscillates from side-to-side. Figure 6 (c) shows the variance of the jet over time with a positive value of computed velocity indicating that the jet has been swept to the right and a negative value indicating that the jet has been swept to the left. It is evident from this plot that the induced electro-vortex flow is unsteady in the presence of convection. This behaviour is different to previous numerical studies of solely electro-vortex flow by Herreman et al. [8] in which the flow was steady and primarily axisymmetric up to very high currents. Therefore, this behaviour must come from the interaction of Rayleigh-Bénard convection and electro-vortex flow and it is very likely that this interaction is the cause of the oscillatory instabilities observed in experimental data by Kelley and Sadoway [28].

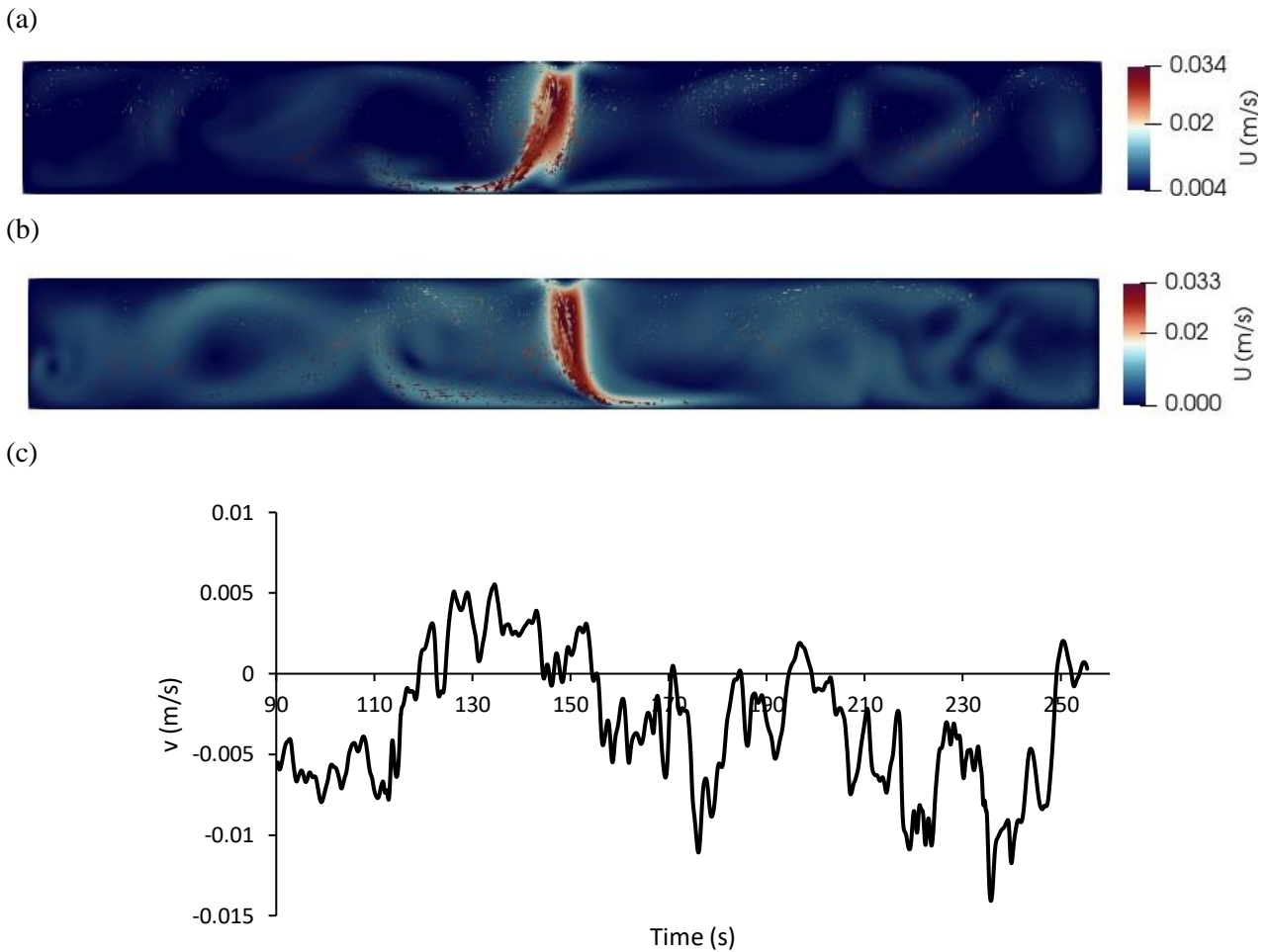

*Figure 6: Convection with 40A of current: instantaneous yz velocity contour (a) at t=205s and (b) at t=135s; (c) instantaneous UY velocity sampled from the computed results under the copper current-collector.*

Comparing the experimental and computed in this work velocity data presented in Figure 7 (a) it can be seen that the shape of the computed velocity profile reflects the experimental data except the peak magnitude of the flow in the central region. It is possible that the difference in velocity magnitude are caused by differences between the external magnetic field measured in Dresden and the local magnetic field inside of the experimental room as was described by Liu et al. [25]. In their numerical model, Liu et al. [25] had to account for fluctuations in the background magnetic field induced by equipment surrounding the experimental apparatus to match the velocities measured in the experiment. This was accomplished by varying the magnitude of the background magnetic field in their simulations. Therefore, the same exercise was undertaken in this work. The vertical component of the background magnetic fields was increased from $B_{\theta z} = 36 \, \mu T$ to $B_{\theta z} = 72 \, \mu T$. The differences in the profile in Figure 7 (a) are considerable. The peak mean velocity increases from ~0.006 m/s to ~0.010 m/s and much more closely reflects the experimentally recorded peak which was ~ 0.012 m/s. The change in magnitude of the background magnetic field between the two cases was small and could easily have been caused by cold working or welding of the stainless steel vessel during construction, resulting in a larger relative permeability [45], or by a horizontal wire.

The increased background vertical magnetic field induces a distinct difference in the velocity streamlines. Rather than the typical vertical streamlines associated with electro-vortex flow, a rotational component can be observed in Figure 7 (b). This rotational movement represents a swirl flow which occurs close to the threshold background magnetic field magnitude predicted by Davidson et al. [44]. Davidson et al. predicted that this would occur with a background magnetic field of a magnitude in the order of 1% of the induced magnetic field. In this work it is found that swirl flow occurs with a background magnetic field closer to 2%. In Davidson et al.'s work the top free-surface has a slip boundary condition while in the present model there was a no-slip condition due to the assumed oxide layer. The additional Ekman layer at the top seems to have a dissipative effect on the rotational component of the induced Lorentz force raising the threshold of background magnetic field required to induce swirl flow.

Swirl flow has already been explored as a means to decrease concentration-overpotentials in liquid metal batteries by Weber et al. [46]. Weber et al. uses a strong vertical magnetic field to drive the flow which results in a decrease in the concentration over-potential in the cathode. The effects of a strong vertical magnetic field on the stability of the battery are unexplored and could have consequences for the onset of the Tayler Instability of Metal Pad Roll. However, the results in this work indicate that a strong vertical magnetic field is not required to obtain a swirl flow, it is likely that swirl flow will already be present in liquid metal batteries from stray vertical magnetic fields.

While the peak velocities recorded in the computational data are much closer to the experimental data with a larger background magnetic field, they do not exactly match. Earlier in this section it was shown that the interaction of Rayleigh-Bénard convection and electro-vortex flow causes horizontal motion in the electro-vortex flow. Therefore, matching the experimental data could be a matter of varying the background magnetic field or, finding boundary conditions for Rayleigh-Bénard convection that better re-create the experiment. However, considering the qualitative agreement with the experimental data this is unnecessary. Further, the computational data from the current numerical model shows much better agreement with the experimental data than that of Ashour et al. [30] also shown in Figure 7 (a). Once again there in the current numerical results there is no upwards vertical velocity induced by joule heating under the copper current collector and the electro-vortex flow has a much higher velocity magnitude than the convection cells. This indicates that electro-vortex flow cannot be neglected as was done by Personettaz et al. [13] and Köllner et al. [12] and that coupled models are required to accurately capture the behaviour of the system.

(a)

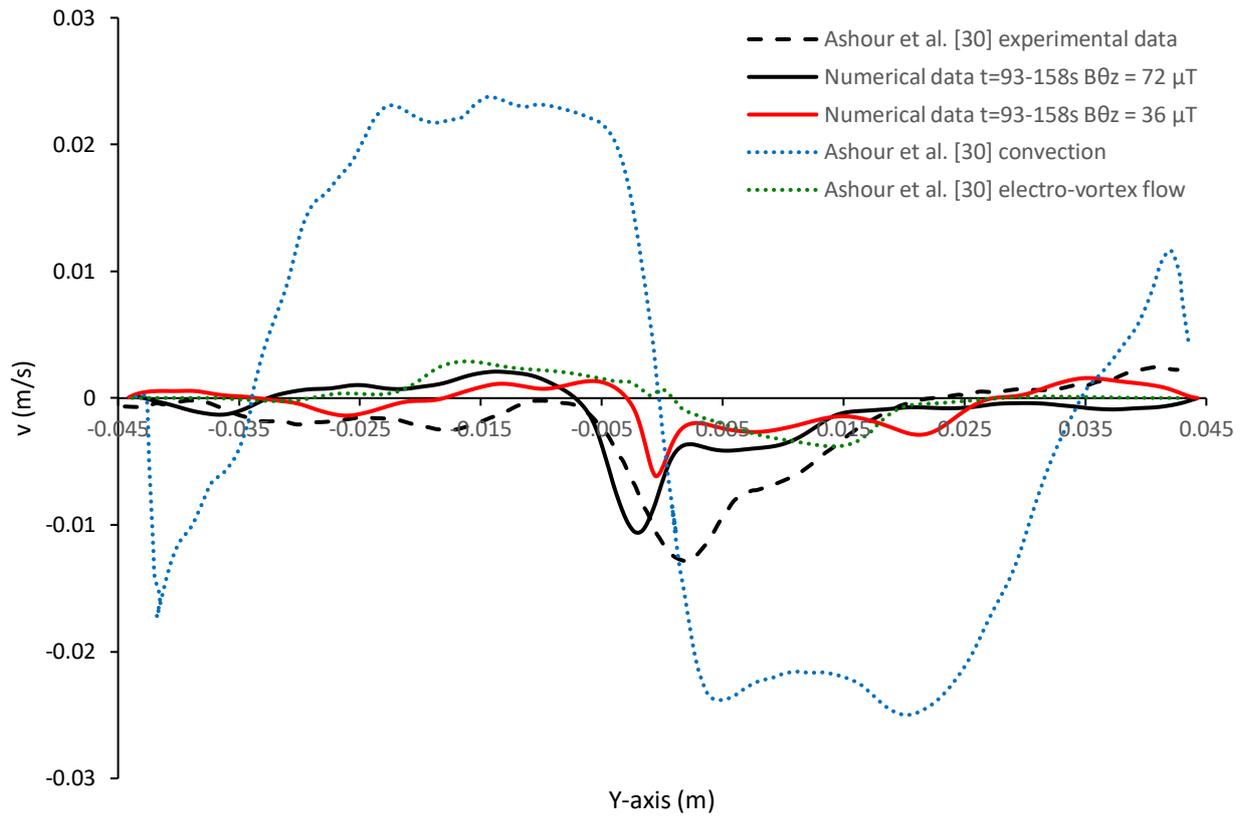

(b)

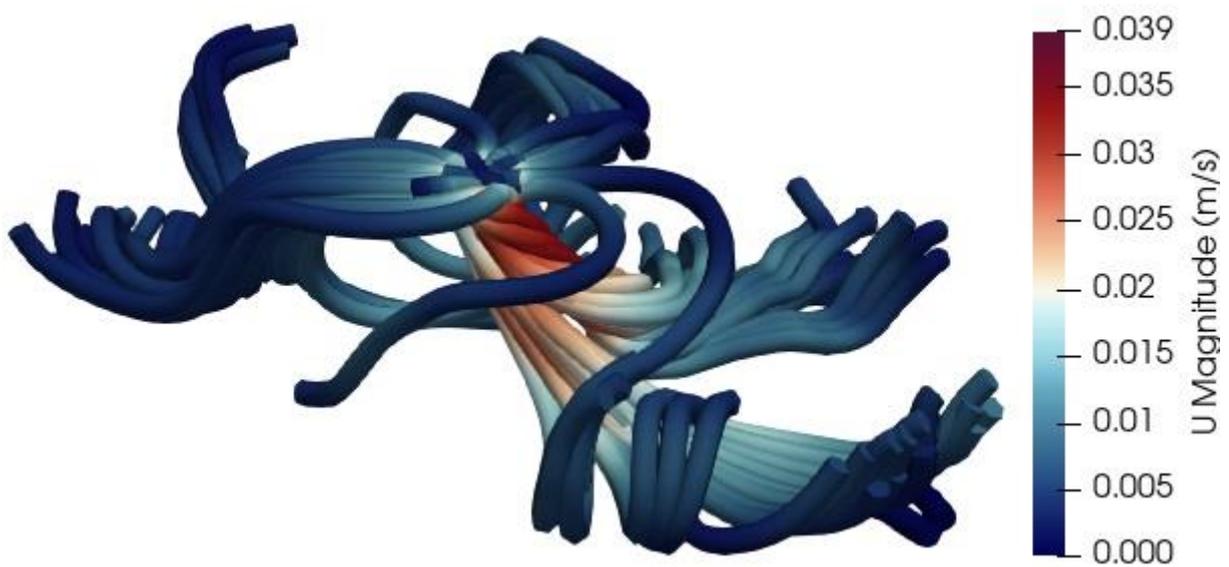

*Figure 7: (a) Comparison of the numerical mean profiles and experimentally obtained profile (b) velocty streamlines of convection with 40 A current and an increased $B_{\theta z} = 72\ \mu T$ background magnetic field*

## 5. Conclusion

A numerical model utilising the low magnetic Reynolds number approximation has been developed and implemented in OpenFOAM. The model has been used to explore the interaction of Rayleigh-Bénard convection and Electro-vortex flow in Liquid Metal Batteries. Convection only in a eutectic Pb-Bi electrode was simulated first and used as the initial condition for the case where 2 A and then 40 A currents were sequentially applied.

It was confirmed that convection in a liquid metal electrode is unsteady and transient. The application of 2A current to the electrode stabilises the flow in good agreement with experimental results however, Rayleigh-Bénard convection was clearly the dominant flow-type. When 40A of current was applied, Electro-vortex flow was found to dominate in the centre of the electrode, not convection, and so it cannot be excluded from future models. Similar to previous works, the background magnetic field was varied to achieve a better match between the experimental and simulation data. In all three cases there was good agreement between the experimental and computed data and the numerical model was found to much better represent the processes occurring in Liquid Metal Batteries than the previously published models.

Variation of the background magnetic field resulted in a swirl flow and it was concluded that this flow would likely be present in all discharging Liquid Metal Batteries. This numerical model can be used to explore the interaction of Electro-vortex flow and Rayleigh-Bénard convection further.

**Acknowledgements**

This work was carried out with the support of an Australian Government Research Training Program (RTP) Scholarship. Computations were performed on the super-computer Gadi with the assistance of resources and services from the National Computational Infrastructure (NCI), which is supported by the Australian Government.